\documentclass[abbrv,aps,prl,twocolumn,showpacs,preprintnumbers,amsmath,amssymb,
superscriptaddress,nofootinbib]{revtex4}

\usepackage{graphicx}% Include figure files
\usepackage{dcolumn}% Align table columns on decimal point
\usepackage{bm}% bold math
\begin{document}
\title{Temperature dependent correlations in covalent insulators}
\author{J. Kune\v{s}}
\email{jan.kunes@physik.uni-augsburg.de}
\affiliation{Theoretical Physics III, Center for Electronic Correlations and Magnetism, Institute of Physics, 
University of Augsburg, Augsburg 
86135, Germany}
\author{V.~I. Anisimov}
\affiliation{Institute of Metal Physics, Russian Academy of
Sciences-Ural Division, 620041 Yekaterinburg GSP-170, Russia}
\date{\today}

\begin{abstract}
Motivated by the peculiar behavior of FeSi and FeSb$_2$ we study the effect of local electronic correlations on magnetic,
transport and optical properties in a specific type of band insulator, namely a covalent insulator. Investigating a
minimum model of covalent insulator within a single-site dynamical mean-field approximation we are able to obtain the crossover
from low temperature non-magnetic insulator to high-temperature paramagnetic metal with parameters
realistic for FeSi and FeSb$_2$ systems. Our results show that the behavior of FeSi does not imply microscopic
description in terms of Kondo insulator (periodic Anderson model) as can be often found in the literature, but in fact reflects generic
properties of a broader class of materials. 
\end{abstract}
\maketitle

The effect of local electronic correlations turning metals into Mott insulators has been one of the major
themes of condensed matter theory. Electronic correlations in band insulators have received much less
attention and were largely limited to investigation of Kondo insulators (KI) described by the periodic Anderson model at half filling.
The possibility of a transition from a band insulator (BI) to a Mott insulator (MI) in the ionic Hubbard model \cite{ionic}
attracted new attention to the correlated BIs.
Among the correlated BIs FeSi \cite{jaccarino}, and more recently FeSb$_2$ \cite{FeSb2}, have been subject of a particular interest
as they are the only known $3d$ materials exhibiting behavior similar to Kondo insulators. 
At low temperatures the electrical resistivity $\rho(T)$ and magnetic susceptibility $\chi(T)$ correspond to thermally activated narrow gap
semiconductor. However, at higher temperatures ($T\geq$300$K$) the slope of $\rho(T)$ changes sign and
becomes positive as in a metal \cite{schlesinger}. The metal-insulator crossover is consistent with the optical
measurements where the gap in the low-$T$ absorption spectrum starts getting filled with increasing temperature, with
no trace of the gap observed \cite{schlesinger,damascelli,PES}
any more at $T\geq$300$K$. 
The magnetic susceptibility $\chi(T)$ vanishes at low $T$, passes through a maximum near 500 $K$, and obeys a Curie-Weiss law at higher temperatures \cite{jaccarino,schlesinger}.
Low electron doping by substituting Fe by Co ($\approx$10\%), or the substitution of Si by Ge ($\approx$30\%)
yield a ferromagnetic metal \cite{Co}. Recently a colossal Seebeck coefficient was reported in FeSb$_2$ \cite{bentien}.

LDA band structure calculations predict FeSi to be a nonmagnetic semiconductor \cite{LDA} in agreement with the low temperature experimental data. 
Interestingly, contrary to the usual underestimation of the gap by LDA, the calculated value $\approx$100meV is 2-3 times
larger than the estimate from optical and photoemission measurements \cite{schlesinger,damascelli,PES}. 
Using the LDA+U approach to mimic the effect of on-site Coulomb repulsion, Anisimov {\it et al.} \cite{Anisimov1} found a
ferromagnetic metallic phase very close by in energy. This approach has also allowed to describe transition from nonmagnetic insulator 
to ferromagnetic metal with substitution Ge for Si in isoelectron electron series FeSi$_{1-x}$Ge$_x$ \cite{Anisimov2},
and the semiconductor-metal transition in FeSb$_2$ \cite{Anisimov3}. Nevertheless, 
current bandstructure theory itself cannot explain the temperature dependence of physical properties.

To describe the susceptibility and other thermodymanic quantities
Mandrus {\it et al.} \cite{mandrus} proposed a temperature independent model density of states  
with two very narrow ($\approx$500$K$) peaks separated by a gap of ($\approx$1000$K$).
The origin of the narrow peaks was attributed to an extreme renormalization of the non-interacting bands similar to Ce-based Kondo insulators invoking
localized states very weakly hybridizing with broad conduction band. However, the assumption of strongly renormalized peaks
surviving up to high temperatures appears rather unlikely. Moreover, the Fe3d-states hybridize very strongly with sp-bands: 
a large contribution of Fe3d-orbitals character can be observed in $\approx$8 eV energy region \cite{LDA,Anisimov3}. 

In this work we study a minimal model of correlated covalent insulator \cite{mermin} within the dynamical-mean field approximation (DMFT). While the construction of the
model is guided by the bandstructure of FeSi and FeSb$_2$, we do not aspire to describe the physics of these materials in its full width.
Our goal is to use simple model free of the complexity of a multi-band low-symmetry system to capture semi-quantitatively the essential physics of FeSi/FeSb$_2$
and provide a well understood reference point for later multi-band study.

\begin{figure}
\includegraphics[angle=270,width=\columnwidth,clip]{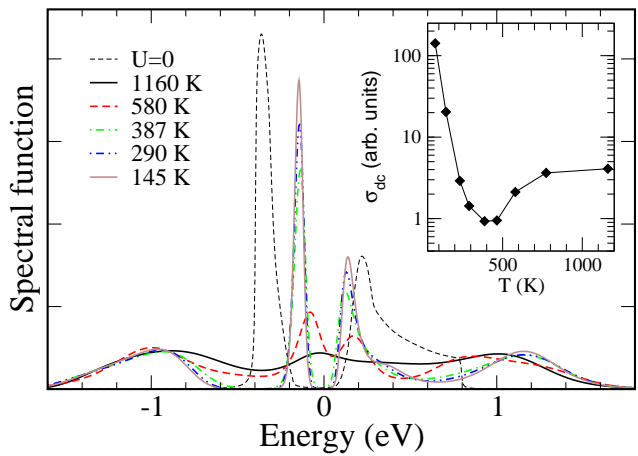}
\caption{\label{fig:spec} (color online) Single-particle spectral density at various temperatures. The non-interacting
spectral function is marked with the dashed-line. The inset shows the temperature
dependence of {\it dc} conductivity.}
\end{figure}
\begin{figure}
\includegraphics[angle=270,width=\columnwidth,clip]{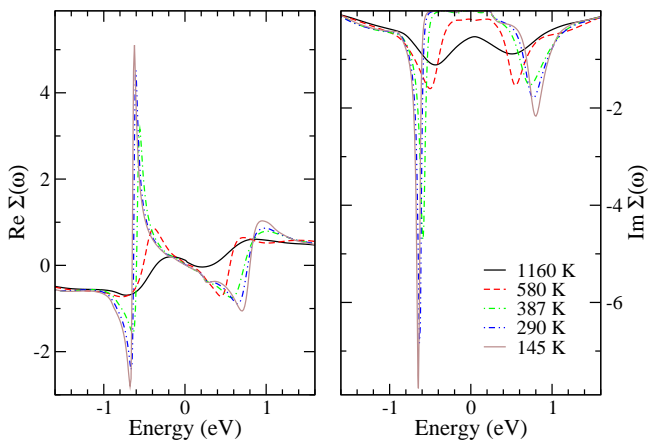}
\caption{\label{fig:Sigma} (color online) Single-particle self-energy $\Sigma(\omega)$ for
various temperatures.}
\end{figure}

In construction of the model we envision Hubbard Hamiltonian on a lattice with one orbital per site, but more than
one site in the primitive cell. The DMFT approximation allows us to calculate important physical quantities, 
in particular the single-particle spectrum, from the knowledge of the non-interacting spectral density  $D(\omega)$ and the on-site repulsion $U$
($U$=1.5 eV throughout this work). The key feature of $D(\omega)$, obtained by truncating the LDA density of states for 
FeSb$_2$ so that the low energy peak and gap structure is captured, is the presence of a gap between bands of the same orbital character.
Fu and Doniach \cite{fu} studied a two-orbital model with an on-site hybridization, which leads to a gapped $D(\omega)$. 
Furmann {\it et al.} \cite{furmann} calculated the single-particle and optical spectra of such model and addressed the
question of interaction driven BI to MI transition at high temperature.  

In general, gapped local spectral function $D(\omega)$ is characteristic of covalent insulator, which is an instance of a band insulator.
In contrast to ionic insulator, arising e.g. from staggered potential or crystal field, with the local orbitals either filled or empty,
the local orbitals of covalent insulator are half (or in general partially) filled. The groundstate of a covalent insulator
is characterized by formation of non-local bonding combinations of orbitals, which are separated by a gap from the anti-bonding 
combinations. Introducing an on-site repulsion $U$ leads to competition between electron localization and formation of non-local bonds.
Large $U$ makes the system a Mott insulator (MI) \cite{moeller}.
The concept of BI to MI transitions driven by the
on-site repulsion was recently studied in the context of the ionic Hubbard model \cite{ionic}.
While analogies can be drawn between the two cases there are also obvious differences 
which make the details BI-MI transitions in the correlated covalent insulator an interesting question.
Here we focus on evolution of physical properties with temperature.

The central quantity in the DMFT framework is the single-particle self-energy $\Sigma(\omega)$ which is determined by the self-consistent 
solution of the equations:
\begin{align}
&G(i\omega_n)=\int d\nu\frac{D(\nu)}{i\omega_n+\mu-\nu-\Sigma(i\omega_n)} \\
&G_0^{-1}(i\omega_n)=G^{-1}(i\omega_n)-\Sigma(i\omega_n),
\end{align}
where $\Sigma[G_0(i\omega_n),U]$ is determined by solving of the auxiliary Anderson impurity problem, for which 
we employ the Hirsch-Fye quantum Monte-Carlo algorithm \cite{qmc}. 
In Fig. \ref{fig:spec} we show the evolution of the single-particle spectral density with temperature $T$. 
At low $T$ renormalized quasiparticle bands with reduced gap (relative to $U=0$ case) and Hubbard bands are well distinguished.
With increasing temperature the lifetime of the quasiparticles decreases (see the self-energy in Fig. \ref{fig:Sigma}) 
and the sharp peaks in the spectral density are smeared as observed
in the photoemission experiment. \cite{PES}
The system thus smoothly evolves from low-$T$ insulator 
to high-$T$ incoherent metal. 

This behavior becomes apparent from the temperature dependence of dc conductivity $\sigma(T)$ calculated according to \cite{cond}
%\begin{equation}
%\label{eq:trans}
\begin{align}
\label{eq:trans}
\sigma(T)\propto\int d\epsilon d\omega \Phi(\epsilon)\rho^2(\epsilon,\omega;T)\frac{\partial f(\omega;T)}{\partial \omega} \\
\rho(\epsilon,\omega;T)=-\frac{1}{\pi} \operatorname{Im}(\omega^+-\epsilon-\Sigma(\omega;T)-\tfrac{i}{\tau})^{-1}\\
\Phi(\epsilon)=\frac{1}{V}\sum_{k}\bigl(\frac{\partial \epsilon_k}{\partial k_x}\bigr)^2\delta(\epsilon-\epsilon_k)\approx
\overline{(\frac{\partial \epsilon_k}{\partial k_x}\bigr)^2}D(\epsilon).
\end{align}
%\end{equation} 
Approximating the squared electron velocity $(\tfrac{\partial \epsilon_k}{\partial k_x})^2$ by its average value
the transport density $\Phi(\epsilon)$ reduces to the spectral density $D(\epsilon)$ times a constant. 
Equation (\ref{eq:trans}) is only valid if the quasiparticle lifetime is finite. To ensure this at all temperatures
we introduce a small constant contribution to the linewidth $1/\tau=0.015$ eV which can be thought of as arising from impurity scattering 
in real materials. Temperature enters formula (\ref{eq:trans}) through the Fermi-Dirac distribution
$f(\omega;T)$ and the self-energy $\Sigma(\omega;T)$. The $T$-dependence of 
resistivity $\rho=\sigma^{-1}$ (see Fig. \ref{fig:spec}) is governed by the Fermi-Dirac distribution at low temperatures
where the thermally activated behavior of band insulator is observed. At higher temperatures the 
increase of the number of thermally activated charge carriers is compensated by the decreasing quasiparticle
lifetime, which eventually becomes dominant and the resistivity assumes the metallic $T$-dependence with positive slope.
This behavior of resistivity is observed in both FeSi \cite{schlesinger} and FeSb$_2$ \cite{FeSb2}.

\begin{figure}
\includegraphics[angle=270,width=\columnwidth,clip]{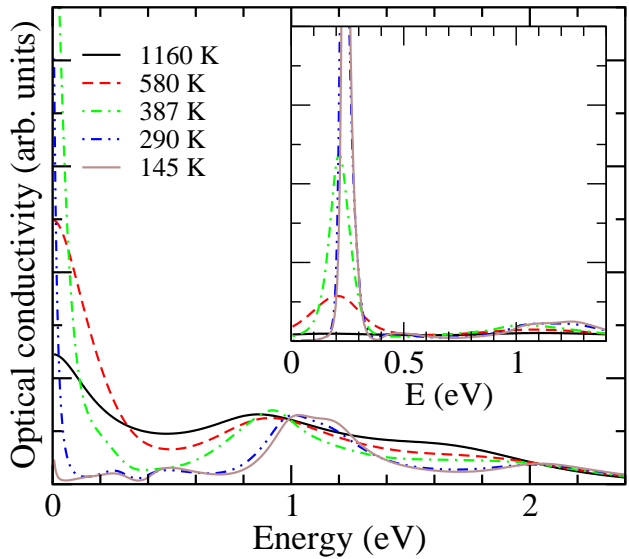}
\caption{\label{fig:opt} (color online) Intra-band contribution to the optical
conductivity at various temperatures. The inset shows the evolution of a
typical inter-band contribution.}
\end{figure}

In general, the conductivity consists of intra-band and inter-band contribution. 
The intra-band contribution is obtained from \cite{cond}
\begin{equation}
\label{eq:opt}
\begin{split}
\sigma(\omega)\propto &\int d\epsilon d\omega' \Phi(\epsilon)\rho(\epsilon,\omega')\rho(\epsilon,\omega'+\omega) \\
&\times\frac{f(\omega')-f(\omega'+\omega)}{\omega}.
\end{split}
\end{equation}
To calculate the inter-band contribution the band dispersion is necessary. In the present case
of a narrow band semiconductor the behavior of the inter-band contribution can be understood by
looking at a typical transition corresponding to a specific $k$-point.
We choose excitation from
the valence band peak at $\epsilon_1$ to conduction band peak at $\epsilon_2$ as the typical transition
\begin{equation}
\label{eq:inter}
\begin{split}
\sigma_{IB}(\omega,k)\propto &M_k\int d\omega' \rho(\epsilon_1,\omega')\rho(\epsilon_2,\omega'+\omega) \\
&\times\frac{f(\omega')-f(\omega'+\omega)}{\omega}.
\end{split}
\end{equation}
The total IB conductivity $\sigma_{IB}(\omega)$ would be obtained by summation over the k-points weighted
by the transition amplitude $M_k$, which results in some broadening due
to the band dispersion. The $T$-dependence of contributions from different $k$-points follows
the same pattern governed by the self-energy $\Sigma(\omega;T)$ and the
Fermi-Dirac function, and thus the typical transition provides a good idea
about the behavior of total $\sigma_{IB}(\omega)$.
The intra-band contribution (\ref{eq:opt}) is shown in
Fig. \ref{fig:opt}. The main effect of decreasing the temperature consists in depletion 
the low frequency region of spectral weight.
Drude peak appears at intermediate temperatures, due to the thermally populated quasiparticle states, 
and disappears at lower temperatures. 
The appearance of the Drude peak is not a generic feature of the model, but depends on the relative
size of the Kondo scale, give by the interaction parameter $U$ and the overall bandwidth,
and the semiconducting gap.
The appearance of well defined quasiparticle states at low temperatures is also reflected
by building-up of a peak in $\sigma_{\text{inter}}$ (inset of Fig. \ref{fig:opt}) above the 
energy of the single-particle gap. Absence of clear Drude peak at high temperatures, and spectral weight
transfer from low frequency region to above the single-particle gap upon cooling  
agrees well with the trend observed experimentally \cite{schlesinger}.

\begin{figure}
\includegraphics[angle=270,width=\columnwidth,clip]{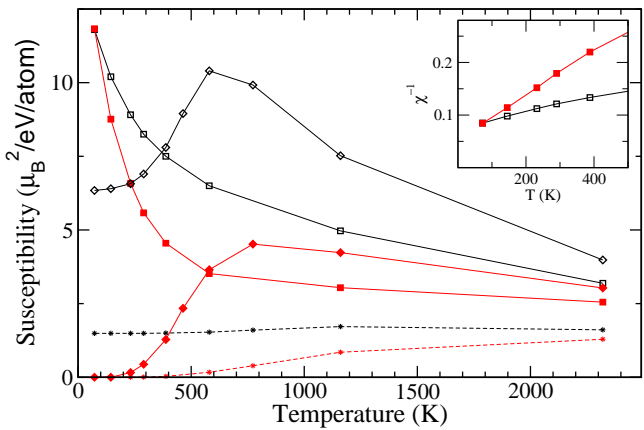}
\caption{\label{fig:chi} (color online) Temperature dependence of local (black) and uniform (red)
 susceptibility for half-filled (diamonds) and doped (squares) system. The dashed lines
  mark the local (black) and uniform (red) susceptibility of the non-interacting system.}
\end{figure}

Next we discuss the behavior of local and uniform magnetic susceptibilities, which are calculated
by integrating the imaginary time correlation function $\langle m_z(\tau)m_z(0)\rangle$ and 
employing the formalism of Ref. \cite{jar} respectively. 
In Fig. \ref{fig:chi} susceptibilities of the system without local interaction are compared
will those of interacting and doped system. At half filling the uniform susceptibility 
exhibits an exponential decrease at low $T$, while the local susceptibility saturates to a finite
value, in both interacting and non-interacting system. The interaction, which favors formation
of local moments, strongly enhances both local and uniform susceptibility and results in appearance
of Curie-Weiss tail at high $T$. Doping of 0.2 electron has a profound effect on the susceptibility.
At this doping level the chemical potential is located close to the valence band peak and thus
a Stoner-like enhancement of the susceptibility can be expected.
While the high $T$ local moment is reduced due to departure from half filling, at low $T$ both the uniform
and local susceptibility show an increase characteristic for local moment systems. 
Different slopes of the inverse local and uniform susceptibilities of the doped system (inset of Fig. \ref{fig:chi})
indicate that a simple Heisenberg model with rigid local moments does 
not provide the appropriate picture. 
% (\emph{magnetic susceptibility diverges with temperature decrease indicating ferromagnetic instability in agreement with experimental ferromagnetic ground state for Fe$_{1-x}$Co$_x$Si} )

Taking the local dynamical correlations into account via DMFT method results
in a satisfactory description of all major FeSi-FeSb$_2$ anomalies. The energy gap present in the single-particle spectrum at low temperatures,
reduced by a factor of two from its LDA value, starts filling with increasing $T$, accompanied by smearing of the sharp peaks,
so that above 600$K$ no trace of the gap is visible.
Corresponding change from insulating to metallic character is observed in the resistivity and the optical conductivity.
The uniform magnetic susceptibility increases exponentially from $T=0$, reaches maximum at $\approx$600$K$ and approaches Curie-Weiss law at higher temperatures.
We find a moderate ($Z\approx 0.5$) very weakly $T$-dependent quasiparticle renormalization below the room temperature implying that
in this temperature range the $T$-dependence of physical quantities such as conductivity or thermopower enters predominantly through the Fermi-Dirac distribution.
This behavior is quite different from classical, e.g. Ce based, Kondo insulators and heavy-fermion compounds where a strong $T$-dependence of the self-energy and
thus the quasiparticle renormalization exists down to very low temperatures.
Sharpening of the spectral features at low temperatures due to quasiparticle renormalization enhances the thermopower, proportional
to the first moment of spectral density \cite{cond}, and thus qualitatively agrees with the experimental observation \cite{bentien}.

In real materials with multiple bands the orbital occupations deviate from half-filling, e.g. due to crystal-fields.
Therefore the effect of competition between covanlency and interaction-driven localization such as
correlation enhancement of susceptibility is expected to be smaller then in the studied model. 
One can also imagine a situation with a staggered potential and a site dependent $U$,
which can then describe materials with strong metal-ligand hybridization such
as LaCoO$_3$. We believe that presented minimal model can provide a useful reference point for understanding
such systems.

In conclusion we have investigated a minimal model of correlated covalent insulator  
to study the unusual temperature evolution of physical properties of FeSi and FeSb$_2$. The model captures all the remarkable properties 
of these materials including the insulator to bad metal transition and the appearance of Curie-Weiss susceptibility at elevated temperatures.

The authors thank D. Vollhardt for discussions and numerous comments.
J.K. gratefully acknowledges the hospitality of the Kavli Institute for Theoretical Physics, Santa Barbara.
This work was supported by the SFB 484 of the Deutsche Forschungsgemeinschaft (J.K., D.V.),
by the Russian Foundation for Basic Research
under the grants  RFFI-07-02-00041 (V.I.A.).

\end{document}